\documentclass[12pt,a4paper]{article}%
\usepackage{epsfig}
\usepackage{amsmath}
\usepackage{amsfonts}
\usepackage{amssymb}
\usepackage{graphicx}
\usepackage[T1]{fontenc}
\usepackage[utf8]{inputenc}
\usepackage{authblk}%
\providecommand{\U}[1]{\protect\rule{.1in}{.1in}}
\begin{document}

\title{Time evolution of quantum systems with time-dependent non-Hermitian
Hamiltonian and the pseudo Hermitian invariant operator}
\author{{\small Mustapha Maamache}$^{a}$\thanks{E-mail: maamache@univ-setif.dz },
{\small Oum Kaltoum Djeghiour}$^{a,b}$\thanks{E-mail:
k.djeghiourjijel@gmail.com}, {\small Walid Koussa}$^{a}$\thanks{E-mail:
koussawalid@yahoo.com} {\small and Naima Mana}$^{a}$\thanks{E-mail:
na3ima\_mn@hotmail.fr}\\$^{(a)}${\small Laboratoire de Physique Quantique et Syst\`{e}mes Dynamiques,}\\{\small Facult\'{e} des Sciences, Universit\'{e} Ferhat Abbas S\'{e}tif 1,
S\'{e}tif 19000, Algeria}\textit{.}\\$^{(b)}${\small D\'{e}partement de Physique, Universit\'{e} de Jijel, \ ,} \\{\small BP 98 \ Ouled Aissa, 18000 Jijel, Algeria.} }
\date{}
\maketitle

\begin{abstract}
We study the time evolution of quantum systems with a time-dependent
non-Hermitian Hamiltonian given by a linear combination of SU(1,1) and SU(2)
generators.With a time-dependent metric, the pseudo-Hermitian invariant
operator is constructed in the same manner as for both the SU(1,1) and SU(2)
systems. The exact common solutions of the Schr\"{o}dinger equations for both
the SU(1,1) and SU(2) systems are obtained in terms of eigenstates of the
pseudo-Hermitian invariant operator.

PACS: 03.65.Ca, 03.65.-w

\end{abstract}

\section{Introduction}

The use of invariants theory to solve quantum systems, whose Hamiltonian is an
explicit function of time, has the advantage to offer an exact solution for
problems solved by the traditional time-dependent perturbation theory. The
existence of invariants (constants of the motion or first integral) introduced
by Lewis \cite{lewis2} and Lewis- Riesenfeld \cite{lewis} is a factor of
central importance in the study of such systems. The invariants method is very
simple due to the relationship between the eigenstates of the invariant
operator and the solutions of the Schr\"{o}dinger equation by means of the
phases; in this case the problem is reduced to find the explicit form of the
invariant operator and the phases. In most cases, use is made of the
Lewis--Riesenfield quadratic invariant to study two archetypal examples. One
of these is the time-dependent generalized harmonic oscillator, the
Hamiltonian of which is a time-dependent function of the SU(1,1) generator and
the other is the spin in a time-dependent varying magnetic field with
Hamiltonian consisting of the SU(2) generator. In \cite{lai1,lai2,cerv,mus1},
the SU(1, 1) and SU(2) time-dependent systems are exactly integrated and the
time evolution operator are obtained thanks to the invariant Hermitian
operator orto the unitary transformation approach.

There is a growing interest in the study of non-Hermitian Hamiltonian
operators due to the fact that these operators may constitute valid quantum
mechanical systems \cite{Scholz,Carl1,Carl5,most1,most3,most4}, because under
certain conditions, non-Hermitian Hamiltonians may have a real spectrum and
therefore may describe realistic physical systems. It has been clarified
\cite{most1,most3,most4} , that a non-Hermitian Hamiltonian having all
eigenvalues real is connected to its Hermitian conjugate through a linear,
Hermitian, invertible and bounded metric operator $\eta=\rho^{+}\rho$ with a
bounded inverse, satisfying \ $H^{+}=\eta H\eta^{-1}$ i.e. $H$ is Hermitian
with respect to a positive definite inner product defined by $\left\langle
.,.\right\rangle _{\eta}=\left\langle .\left\vert \eta\right\vert
.\right\rangle $ and called as $\eta$ -pseudo-Hermitian. Essentially the same
idea had appeared previously under the name of \textquotedblleft
quasi-Hermiticity\textquotedblright\ by Scholtz et al \cite{Scholz}. It is
also established \cite{most1,most3,most4} that the non Hermitian Hamiltonian
$H$ can be transformed to an equivalent Hermitian one given by $h=\rho
H\rho^{-1}$, where $h$ is the equivalent Hermitian analog of $H$ with respect
to the standard inner product $\left\langle .,.\right\rangle .$

All these efforts have been devoted to study time-independent non-Hermitian
systems. Whereas the treatment for systems with time-dependent non-Hermitian
Hamiltonians with time-independent metric operators have been extensively
studied \cite{Faria1,Faria2}, the generalization to time-dependent metric
operators is quite controversial
\cite{znojil1,most5,most6,znojil2,znojil3,Bila,wang1,wang2,mus,fring1,fring2,khant,frith,luiz1,luiz2}%
.

Recent contributions \cite{fring1,fring2} have advanced the grounds for
treating time-dependent non-Hermitian Hamiltonians through time dependent
Dyson maps. It has been argued that it is incompatible to maintain unitary
time evolution for time-dependent non-Hermitian Hamiltonians when the metric
operator is explicitly time dependent.

In light of the above discussion, one important question motivates our work
here: How can we treat a non-Hermitian SU(1, 1) and SU(2) time-dependent
quantum problems and investigate the possibility of finding the exact solution
of the Schr\"{o}dingerr equation in terms of eigenstates of the pseudo-
invariant operator as well the real\ associated phases?

Let's first briefly recall the pseudo-Hermitian invariants theory
\cite{khant,mus2}. The invariant operator $I^{PH}(t)$ is said to be
pseudo-Hermitian with respect to $\eta(t)$ if%
\begin{equation}
I^{PH\dag}\left(  t\right)  =\eta(t)I^{PH}\left(  t\right)  \eta^{-1}(t)\text{
}\Leftrightarrow I^{h}(t)=\rho(t)I^{PH}(t)\rho^{-1}(t)=I^{h\dag}%
(t),\label{quasi}%
\end{equation}
where $\eta(t)=\rho^{+}(t)\rho(t)$ is a linear time dependent Hermitian
invertible operator. Thus $I^{PH}\left(  t\right)  $ may be mapped to the
Hermitian invariant operator $I^{h}\left(  t\right)  $, by a similarity
transformation $\rho(t)$.

The solutions of the time-dependent Schr\"{o}dinger equation ( $\hbar=c=1$ are
used throughout)
\begin{equation}
i\frac{\partial}{\partial t}\left\vert \Phi^{H}(t)\right\rangle
=H(t)\left\vert \Phi^{H}(t)\right\rangle , \label{shro}%
\end{equation}
for the non- Hermitian Hamiltonian $H(t)$ can be found with the aid of the
quantum pseudo invariant method of Lewis and Riesenfeld. A pseudo-Hermitian
invariant operator $I^{PH}(t)$ for a given non-Hermitian Hamiltonian $H(t)$ is
defined to satisfy%

\begin{equation}
\frac{\partial I^{PH}(t)}{\partial t}=i\left[  I^{PH}\left(  t\right)
,H(t)\right]  , \label{lewisPH}%
\end{equation}
where $I^{PH}\left(  t\right)  $ has a finite number of nondegenerate
eigenstates $\left\vert \phi_{n}^{H}(t)\right\rangle $ satisfying%

\begin{equation}
\text{ \ }I^{PH}\left(  t\right)  \left\vert \phi_{n}^{H}(t)\right\rangle
=\lambda_{n}\left\vert \phi_{n}^{H}(t)\right\rangle ,
\end{equation}
and\
\begin{equation}
\left\langle \phi_{m}^{H}(t)\right\vert \eta(t)\left\vert \phi_{n}%
^{H}(t)\right\rangle =\delta_{m,n}%
\end{equation}
with time-independent eigenvalues $\lambda_{n}$ . Since the Hermitian
invariant $I^{h}(t)$ and the non-Hermitian invariant $I^{PH}(t)$ are related
by a similarity transformation (\ref{quasi}), they belong to the same
similarity class and therefore have the same eigenvalues. The reality of the
eigenvalues $\lambda_{n}$ is guaranteed, since one of the invariants involved,
i.e. $I^{h}(t),$ is Hermitian.\ 

\bigskip Now, if the exact invariant $I^{PH}\left(  t\right)  $ (constant of
motion) exists and does not contain any time derivative operators, we can
write the solutions of the Schr\"{o}dinger equation (\ref{shro}) in terms of
the eigenfunctions $\left\vert \phi_{n}^{H}(t)\right\rangle $ of
$I^{PH}\left(  t\right)  $,%
\begin{equation}
\left\vert \Phi_{n}^{H}(t)\right\rangle =e^{i\varphi_{n}(t)}\left\vert
\phi_{n}^{H}(t)\right\rangle , \label{sol}%
\end{equation}
the phase functions $\varphi_{n}(t)$ are derived from the equation:%
\begin{equation}
\frac{d\varphi_{n}(t)}{dt}=\left\langle \phi_{n}^{H}(t)\right\vert
\eta(t)\left[  i\hbar\frac{\partial}{\partial t}-H(t)\right]  \text{\ }%
\left\vert \phi_{n}^{H}(t)\right\rangle . \label{Phase}%
\end{equation}
In Eq. (\ref{Phase}), the first term is parallel to a familiar non-adiabatic
geometrical phase, but the second term$,$ representing effects due to a
time-dependent Hamiltonian, is a dynamical phase. The sum of these two terms
that can ensure that the phase functions $\varphi_{n}(t)$ are real.

The general solution of the Schr\"{o}dinger equation for the system with
non-Hermitian time-dependent Hamiltonians $H(t)$ and pseudo-Hermtian invariant
operator are readily obtained as follows:%

\begin{equation}
\left\vert \Phi^{H}(t)\right\rangle =%
{\textstyle\sum_{n}}
C_{n}e^{i\alpha_{n}(t)}\left\vert \phi_{n}^{H}(t)\right\rangle \label{Gensol}%
\end{equation}
where the $C_{n}$ = $\left\langle \phi_{n}^{H}(0)\right\vert \eta(0)\left\vert
\Phi^{H}(0)\right\rangle $ are time-independent coefficients.

Let's note that, it then follows immediately by direct substitution of
(\ref{quasi}) into (\ref{lewisPH}) that the two invariants operators
$I^{PH\dag}\left(  t\right)  $ and $I^{h}(t)$ satisfy the following equations:%

\begin{equation}
\frac{\partial I^{PH+}(t)}{\partial t}=i\left[  I^{PH\dag}\left(  t\right)
,\eta\left(  t\right)  H\left(  t\right)  \eta^{-1}\left(  t\right)
+i\dot{\eta}\left(  t\right)  \eta^{-1}\left(  t\right)  \right]  ,
\end{equation}

\begin{equation}
\frac{\partial I^{h}(t)}{\partial t}=i\left[  I^{h}\left(  t\right)
,\rho\left(  t\right)  H\left(  t\right)  \rho^{-1}\left(  t\right)
+i\dot{\rho}\left(  t\right)  \rho^{-1}\left(  t\right)  \right]  ,
\end{equation}
which show that the non-Hermitian Hamiltonian $H(t)$ is related to its
Hermitian conjugate $H^{\dag}\left(  t\right)  $ as
\begin{equation}
H^{\dag}\left(  t\right)  =\eta\left(  t\right)  H\left(  t\right)  \eta
^{-1}\left(  t\right)  +i\dot{\eta}\left(  t\right)  \eta^{-1}\left(
t\right)  , \label{PHH1}%
\end{equation}
\ and the Hermitian Hamiltonian $h(t)$ is linked to the non-Hermitian
Hamiltonian $H\left(  t\right)  $ by the time-dependent Dyson equation%
\begin{equation}
h\left(  t\right)  =\rho\left(  t\right)  H\left(  t\right)  \rho^{-1}\left(
t\right)  +i\dot{\rho}\left(  t\right)  \rho^{-1}\left(  t\right)  ,
\label{PHH2}%
\end{equation}
The above equations have been obtained by Fring and Moussa
\cite{fring1,fring2} by assuming that the two solutions $\left\vert \Phi
^{H}(t)\right\rangle $ and $\left\vert \Psi^{h}(t)\right\rangle $ of the two
time-dependent Schrodinger equations ruled by $H\left(  t\right)  $ and
$h\left(  t\right)  $ respectively, are related by a time-dependent invertible
operator $\eta(t)$ as $\left\vert \Psi^{h}(t)\right\rangle =\eta(t)\left\vert
\Phi^{H}(t)\right\rangle $. Then, they argued that the time-dependent
quasi-Hermiticity relation and the time-dependent Dyson equation can be solved
consistently in such scenario for a time-dependent Dyson map and
time-dependent metric operator, respectively.

Our approach is different from Fring's and Moussa's one, because we resolve
the standard quasi-Hermiticity relation and the standard Dyson equation
(\ref{quasi}) for a time-dependent invariant operator with time-dependent
$\eta(t)$ and a time-dependent similarity transformation $\rho(t).$ While the
key feature in Fring's and Moussa's approach is that the relation (\ref{PHH1})
is stated as the time-dependent quasi-Hermiticity relation. We believe that
the resolution of the time-dependent Dyson equation and the time-dependent
quasi-Hermiticity relation stated by Fring and Moussa become more difficult
due to the presence of the last term in equations (\ref{PHH1}) and (\ref{PHH2}).

In this\ paper, we answer this question from a new perspective by studying the
time-dependent non-Hermitian Hamiltonian systems given by a linear combination
of SU(1, 1) and SU(2) generators using a pseudo-invariant operator theory
which is constructed in a manner as for both the SU(1, 1) and SU(2) systems.
An advantage of the pseudo- invariant operator is that it allows to obtain the
exact solution of the Schr\"{o}dinger equation in terms of eigenstates of the
invariant operator as well as the time-evolution operator.

\section{Evolution of non-Hermitian SU(1, 1) and SU(2) time-dependent systems}

The SU(1, 1) and SU(2) time-dependent systems that we consider are described
by the non-Hermitian Hamiltonian%
\begin{equation}
H(t)=2\omega(t)K_{0}+2\alpha(t)K_{-}+2\beta(t)K_{+}, \label{HH}%
\end{equation}
where $\left(  \omega(t),\alpha(t),\beta(t)\right)  $ $\in C$ are arbitrary
functions of time. $K_{0}$ is a Hermitian operator, while $K_{+}=\left(
K_{-}\right)  ^{+}$. The commutation relations between these operators are%

\begin{equation}
\left\{
\begin{array}
[c]{c}%
\left.  \left[  K_{0},K_{+}\right]  =K_{+}\right. \\
\left.  \left[  K_{0},K_{-}\right]  =-K_{-}\right. \\
\left.  \left[  K_{+},K_{-}\right]  =DK_{0}\right.
\end{array}
\right.  . \label{lie}%
\end{equation}
The Lie algebra of SU(1, 1) and SU(2) consists of the generators $K_{0}$,
$K_{-}$ and $K_{+}$ corresponding to $D=-2$ \ and $2$ in the commutation
relations (\ref{lie}), respectively.

\bigskip In what follows, we investigate the quantum dynamics of our
time-dependent systems (\ref{shro}) associated with the Hamiltonian (\ref{HH})
. To this end, we consider the most general invariant $I^{PH}(t)$ in the form
\begin{equation}
I^{PH}(t)=2\delta_{1}(t)K_{0}+2\delta_{2}(t)K_{-}+2\delta_{3}(t)K_{+},
\label{IH}%
\end{equation}
where $\delta_{1}(t)$, $\delta_{2}(t)$, $\delta_{3}(t)$ are time dependent
real parameters$.$ The invariant (\ref{IH}) is of course manifestly
non-Hermitian when $\delta_{2}(t)\neq$ $\delta_{3}(t).$

As is well known \cite{cheng,kli,bar} an element of the group of SU(1, 1) or
SU(2) can be obtained by exponentiation of an element of the corresponding
algebra. It is also well known that we can write down this element in many
equivalent factorized ways. The Baker--Hausdorff--Campbell formula allows us
to express all elements of SU(1, 1) or of \ SU(2) obtained by exponentiation
of an Hermitic element of SU(1, 1) or of SU(2) as%

\begin{align}
\rho\left(  t\right)   &  =\exp\left\{  2\left[  \epsilon\left(  t\right)
K_{0}+\mu\left(  t\right)  K_{-}+\mu^{\ast}\left(  t\right)  K_{+}\right]
\right\}  ,\nonumber\\
&  =\exp\left[  \vartheta_{+}\left(  t\right)  K_{+}\right]  \exp\left[
\ln\vartheta_{0}\left(  t\right)  K_{0}\right]  \exp\left[  \vartheta
_{-}\left(  t\right)  K_{-}\right]  , \label{metr}%
\end{align}
where
\begin{align}
\vartheta_{+}\left(  t\right)   &  =\frac{2\mu^{\ast}\sinh\theta}{\theta
\cosh\theta-\epsilon\sinh\theta}=-\zeta(t)e^{-i\varphi(t)},\nonumber\\
\vartheta_{0}\left(  t\right)   &  =\left(  \cosh\theta-\frac{\epsilon}%
{\theta}\sinh\theta\right)  ^{-2}=-\frac{D}{2}\zeta^{2}(t)-\chi(t),
\label{TDC}\\
\vartheta_{-}\left(  t\right)   &  =\frac{2\mu\sinh\theta}{\theta\cosh
\theta-\epsilon\sinh\theta}=-\zeta(t)e^{i\varphi(t)},\nonumber\\
\chi(t)  &  =-\frac{\cosh\theta+\frac{\epsilon}{\theta}\sinh\theta}%
{\cosh\theta-\frac{\epsilon}{\theta}\sinh\theta}\text{ \ \ \ \ \ \ ,\ \ \ }%
\theta=\sqrt{\epsilon^{2}+2D\left\vert \mu\right\vert ^{2}}.\nonumber
\end{align}
\ This factorization is valid for SU(1, 1) ($D=-2$) and for SU(2) ($D=2$).

The key point of our method is to solve the standard quasi-Hermiticity
relation (\ref{quasi}) by making, for simplicity, the Hermitian ansatz
(\ref{metr}) for time-dependent invertible operator $\rho\left(  t\right)  $
\ Let us solve the standard quasi-Hermiticity relation (\ref{quasi}) by making
the following general and, for simplicity, Hermitian ansatz for a time
dependent metric$\ \rho\left(  t\right)  .$We obtain, after some algebra, the
transformed invariant operator $I^{h}(t)=\rho(t)I^{PH}(t)\rho^{-1}(t)$
\begin{align}
I^{h}(t)  &  =\frac{2}{\vartheta_{0}}\left[  \left[  \left(  \frac{D}%
{2}\vartheta_{-}\vartheta_{+}-\chi\right)  \delta_{1}+D\left(  \vartheta
_{+}\delta_{2}+\chi\vartheta_{-}\delta_{3}\right)  \right]  K_{0}\right.
\nonumber\\
&  \left.  +\left(  \vartheta_{-}\delta_{1}+\delta_{2}-\frac{D}{2}%
\vartheta_{-}^{2}\delta_{3}\right)  K_{-}+\left(  \chi\vartheta_{+}\delta
_{1}-\frac{D}{2}\vartheta_{+}^{2}\delta_{2}+\chi^{2}\delta_{3}\right)
K_{+}\right]  . \label{invh}%
\end{align}
The derivation of equation (\ref{invh}) is made of the following identities:
\begin{equation}
\left\{
\begin{array}
[c]{c}%
\exp\left[  \vartheta_{-}K_{-}\right]  K_{0}\exp\left[  -\vartheta_{-}%
K_{-}\right]  =K_{0}+\vartheta_{-}K_{-}\\
\exp\left[  \vartheta_{+}K_{+}\right]  K_{0}\exp\left[  -\vartheta_{+}%
K_{+}\right]  =K_{0}-\vartheta_{+}K_{+}%
\end{array}
\right.  ,
\end{equation}

\begin{equation}
\left\{
\begin{array}
[c]{c}%
\exp\left[  \ln\vartheta_{0}K_{0}\right]  K_{-}\exp\left[  -\ln\vartheta
_{0}K_{0}\right]  =\frac{K_{-}}{\vartheta_{0}}\\
\exp\left[  \vartheta_{+}K_{+}\right]  K_{-}\exp\left[  -\vartheta_{+}%
K_{+}\right]  =K_{-}+D\vartheta_{+}K_{0}-\frac{D}{2}\vartheta_{+}^{2}K_{+}%
\end{array}
\right.  ,
\end{equation}

\begin{equation}
\left\{
\begin{array}
[c]{c}%
\exp\left[  \ln\vartheta_{0}K_{0}\right]  K_{+}\exp\left[  -\ln\vartheta
_{0}K_{0}\right]  =\vartheta_{0}K_{+}\\
\exp\left[  \vartheta_{-}K_{-}\right]  K_{+}\exp\left[  \vartheta_{-}%
K_{-}\right]  =K_{+}-D\vartheta_{-}K_{0}-\frac{D}{2}\vartheta_{-}^{2}K_{-}%
\end{array}
\right.  .
\end{equation}
For $I^{h}(t)$ to be Hermitian ($I^{h}(t)=I^{+h}(t)$) we require the
coefficient of $K_{0}$ is real, and the coefficients of $K_{-}$ and $K_{+}$
are complex conjugate of one another. Using these two requirements, we have:%

\begin{align}
\left[  \left(  \frac{D}{2}\vartheta_{-}\vartheta_{+}-\chi\right)  \delta
_{1}+D\left(  \vartheta_{+}\delta_{2}+\chi\vartheta_{-}\delta_{3}\right)
\right]   &  =\left[  \left(  \frac{D}{2}\vartheta_{-}\vartheta_{+}%
-\chi\right)  \delta_{1}+D\left(  \vartheta_{-}\delta_{2}+\chi\vartheta
_{+}\delta_{3}\right)  \right]  ,\nonumber\\
\left(  \vartheta_{-}\delta_{1}+\delta_{2}-\frac{D}{2}\vartheta_{-}^{2}%
\delta_{3}\right)   &  =\left(  \chi\vartheta_{-}\delta_{1}-\frac{D}%
{2}\vartheta_{-}^{2}\delta_{2}+\chi^{2}\delta_{3}\right)  ,\\
\left(  \chi\vartheta_{+}\delta_{1}-\frac{D}{2}\vartheta_{+}^{2}\delta
_{2}+\chi^{2}\delta_{3}\right)   &  =\left(  \vartheta_{+}\delta_{1}%
+\delta_{2}-\frac{D}{2}\vartheta_{+}^{2}\delta_{3}\right)  ,\nonumber
\end{align}
from the first constraint we derive the equality
\begin{equation}
\delta_{2}=\delta_{3}\chi, \label{delta1}%
\end{equation}
while the other two constraints leads to%

\begin{align}
\delta_{1}  &  =\frac{\left(  \frac{D}{2}\vartheta_{-}^{2}-\chi\right)
}{\vartheta_{-}}\delta_{3},\nonumber\\
\delta_{1}  &  =\frac{\left(  \frac{D}{2}\vartheta_{+}^{2}-\chi\right)
}{\vartheta_{+}}\delta_{3}. \label{delta}%
\end{align}
From the equations (\ref{delta}), it follows that $\vartheta_{+}(t)=$
$\vartheta_{-}(t)\equiv-\zeta(t)$ implying that the time dependent parameter
$\mu(t)$ must be real, i.e. $\mu(t)=\mu^{\ast}(t)$. Finally the similarity
transformation\ (\ref{metr}) maps the non-Hermitian quadratic invariant
\ (\ref{IH}) into $I^{h}(t)$ given by%

\begin{equation}
I^{h}\left(  t\right)  =\frac{2}{\vartheta_{0}}\left[  \left(  \frac{D}%
{2}\zeta^{2}-\chi\right)  \delta_{1}-2D\chi\zeta\delta_{3}\right]  K_{0}.
\label{invh1}%
\end{equation}

Let $\left\vert \psi_{n}^{h}\right\rangle $ be the eigenstate of $K_{0}$ with
eigenvalue $k_{n}$ i.e.%
\begin{equation}
K_{0}\left\vert \psi_{n}^{h}\right\rangle =k_{n}\left\vert \psi_{n}%
^{h}\right\rangle .
\end{equation}
The eigenstates of $I^{h}\left(  t\right)  $ (\ref{invh1}) are obviously given
by
\begin{equation}
I^{h}\left(  t\right)  \left\vert \psi_{n}^{h}(t)\right\rangle =\frac
{2}{\vartheta_{0}}\left[  \left(  \frac{D}{2}\zeta^{2}-\chi\right)  \delta
_{1}-2D\chi\zeta\delta_{3}\right]  k_{n}\left\vert \psi_{n}^{h}\right\rangle
\text{, \ }%
\end{equation}
because of the time-dependence, the invariant $I^{h}\left(  t\right)  $ is a
conserved quantity whose eigenvalues are real constants. However, without loss
of generality, the factor $\left[  \left(  D\zeta^{2}/2-\chi\right)
\delta_{1}-2D\chi\zeta\delta_{3}\right]  /\vartheta_{0}$ can be taken equal to
$1.$\ It follows that the eigenstate $\left\vert \phi_{n}^{H}(t)\right\rangle
$ of $I^{PH}(t)$ can be directly deduced from the basis $\left\vert \psi
_{n}^{h}\right\rangle $ of its Hermitian counterpart $I^{h}\left(  t\right)  $
through the similarity transformation $\left\vert \phi_{n}^{H}(t)\right\rangle
=$ $\rho^{-1}(t)\left\vert \psi_{n}^{h}\right\rangle $ with time-independent
eigenvalue $k_{n}.$

According to the above discussions, the problem is reduced to find a pseudo
Hermitian invariant operator and the suitable real phases of its
eigenfunctions to take them as a\ solution for the Schr\"{o}dinger equation.
In a first step, we will determine the real parameters $\delta_{1},\delta
_{2},\delta_{3}$ so that our invariant operator $I^{PH}(t)$ (\ref{IH}) is
pseudo Hermitian. Imposing the quasi- Hermiticity condition (\ref{quasi}) on
$I^{h}\left(  t\right)  ,$ we get%

\begin{align}
I^{\dag PH}(t)  &  =\rho^{+}\left(  t\right)  I^{h}\left(  t\right)
\rho^{-1+}\left(  t\right)  =2\delta_{1}\hat{K}_{0}+2\delta_{3}\hat{K}%
_{-}+2\delta_{2}\hat{K}_{+}\nonumber\\
&  =\frac{2}{\vartheta_{0}}\left[  \left(  \frac{D}{2}\zeta^{2}-\chi\right)
\hat{K}_{0}-\zeta\hat{K}_{-}-\chi\zeta\hat{K}_{+}\right]  .
\end{align}
From the above equation the real parameters $\delta_{1},\delta_{2},\delta_{3}$
follow straightforwardly:%
\begin{equation}
\delta_{1}=\frac{\left(  \frac{D}{2}\zeta^{2}-\chi\right)  }{\vartheta_{0}%
}\text{ , }\delta_{2}=-\frac{\chi\zeta}{\vartheta_{0}}\text{ , }\delta
_{3}=-\frac{\zeta}{\vartheta_{0}}.
\end{equation}
Therefore, the pseudo Hermitian invariant operator $I^{PH}(t)$ is written in
the following form%

\begin{equation}
I^{PH}(t)=\frac{2}{\vartheta_{0}}\left[  \left(  \frac{D}{2}\zeta^{2}%
-\chi\right)  K_{0}-\chi\zeta K_{-}-\zeta K_{+}\right]  . \label{PH1}%
\end{equation}

The second step in the method is imposing for $I^{PH}(t)$(\ref{PH1}) the
invariance condition (\ref{lewisPH}) which lead to the following relations :%

\begin{equation}
\dot{\vartheta}_{0}=\frac{2\vartheta_{0}}{\zeta}\left[  -2\zeta\left\vert
\omega\right\vert \sin\varphi_{\omega}+\left\vert \alpha\right\vert
\sin\varphi_{\alpha}+\left(  \chi-D\zeta^{2}\right)  \left\vert \beta
\right\vert \sin\varphi_{\beta}\right]  , \label{cont1}%
\end{equation}

\begin{equation}
\overset{\cdot}{\zeta}=-2\zeta\left\vert \omega\right\vert \sin\varphi
_{\omega}+2\left\vert \alpha\right\vert \sin\varphi_{\alpha}-D\zeta
^{2}\left\vert \beta\right\vert \sin\varphi_{\beta}, \label{cont2}%
\end{equation}

\begin{equation}
\
\begin{array}
[c]{c}%
\chi\left\vert \beta\right\vert \cos\varphi_{\beta}=\left\vert \alpha
\right\vert \cos\varphi_{\alpha}\\
\left(  \chi-\frac{D}{2}\zeta^{2}\right)  \left\vert \alpha\right\vert
\cos\varphi_{\alpha}=\chi\zeta\left\vert \omega\right\vert \cos\varphi
_{\omega}\\
\zeta\left\vert \omega\right\vert \cos\varphi_{\omega}=\left(  \chi-\frac
{D}{2}\zeta^{2}\right)  \left\vert \beta\right\vert \cos\varphi_{\beta}%
\end{array}
, \label{rel}%
\end{equation}
here, $\varphi_{\omega}$, $\varphi_{\alpha}$ , and $\varphi_{\beta}$ are the
polar angles of $\omega$, $\alpha$, and $\beta$, respectively.

The final step consists in determining the Schrodinger solution (\ref{sol})
which is an eigenstate of the pseudo Hermitian invariant (\ref{PH1})
multiplied by a time-dependent factor (\ref{Phase})%

\begin{align}
\frac{d\varphi_{n}(t)}{dt}  &  =\left\langle \phi_{n}^{H}(t)\right\vert
\eta(t)\left[  i\frac{\partial}{\partial t}-H(t)\right]  \text{\ }\left\vert
\phi_{n}^{H}(t)\right\rangle \nonumber\\
&  =\left\langle \psi_{n}^{h}\right\vert \left[  i\rho\dot{\rho}^{-1}-\rho
H\rho^{-1}\right]  \text{\ }\left\vert \psi_{n}^{h}\right\rangle .
\label{Phase1}%
\end{align}
Using the non-Hermitian Hamiltonian $H(t)$ (\ref{HH}) and then deriving the
transformed Hamiltonian $\left[  i\rho\dot{\rho}^{-1}-\rho H\rho^{-1}\right]
$\ through the metric operator $\rho(t)$ (\ref{metr}), we further identify
this transformed Hamiltonian as
\begin{equation}
i\rho\dot{\rho}^{-1}-\rho H\rho^{-1}=2W\left(  t\right)  K_{0}+2U\left(
t\right)  K_{-}+2V\left(  t\right)  K_{+},
\end{equation}
where the coefficient functions are%
\begin{align}
W\left(  t\right)   &  =-\frac{1}{\vartheta_{0}}\left[  \omega\left(  \frac
{D}{2}\zeta^{2}-\chi\right)  -D\zeta\left(  \alpha+\beta\chi\right)  +\frac
{i}{2}\left(  \dot{\vartheta}_{0}+D\zeta\overset{\cdot}{\zeta}\right)
\right]  ,\\
U\left(  t\right)   &  =\frac{1}{\vartheta_{0}}\left[  \omega\zeta
-\alpha+\frac{D}{2}\beta\zeta^{2}+i\frac{\overset{\cdot}{\zeta}}{2}\right]
,\\
V\left(  t\right)   &  =\frac{1}{\vartheta_{0}}\left[  \omega\chi\zeta
+\frac{D}{2}\alpha\zeta^{2}-\beta\chi^{2}-\frac{i}{2}\left(  \zeta
\dot{\vartheta}_{0}-\vartheta_{0}\overset{\cdot}{\zeta}+\frac{D}{2}%
\vartheta^{2}\overset{\cdot}{\zeta}\right)  \right]  .
\end{align}
By using Eqs. (\ref{rel}), the above time-dependent coefficients $U$ $,V$ are
identically equal to zero ($U$ $=V=0$), whereas the coefficients $W$ is
reduced to
\begin{equation}
W\left(  t\right)  =-\frac{1}{\vartheta_{0}}\left\{  \left(  \frac{D}{2}%
\zeta^{2}-\chi\right)  \left\vert \omega\right\vert \cos\varphi_{\omega
}-2D\zeta\left\vert \alpha\right\vert \cos\varphi_{\alpha}-i\frac
{\vartheta_{0}}{\zeta}\left[  \zeta\left\vert \omega\right\vert \sin
\varphi_{\omega}-\left\vert \alpha\right\vert \sin\varphi_{\alpha}%
-\chi\left\vert \beta\right\vert \cos\varphi_{\beta}\right]  \right\}  .
\end{equation}
Knowing that the phase $\varphi_{n}(t)$ (\ref{Phase1}) must be real, we need
to impose that the frequency $W\left(  t\right)  $ is real. Then, we obtain
the exact phase of the eigenstate%

\begin{equation}
\varphi_{n}(t)=-2k_{n}%
{\displaystyle\int\limits_{0}^{t}}
\frac{1}{\vartheta_{0}}\left[  \left(  \frac{D}{2}\zeta^{2}-\chi\right)
\left\vert \omega\right\vert \cos\varphi_{\omega}-2D\zeta\left\vert
\alpha\right\vert \cos\varphi_{\alpha}\right]  dt^{\prime}. \label{phase1}%
\end{equation}
Therefore, the general solution (\ref{Gensol})of the Schr\"{o}dinger equation
is given by
\begin{equation}
\left\vert \Phi^{H}(t)\right\rangle =%
{\textstyle\sum_{n}}
C_{n}(0)\exp\left(  -ik_{n}%
{\displaystyle\int\limits_{0}^{t}}
\frac{2}{\vartheta_{0}}\left[  \left\vert \omega\right\vert \left(  \frac
{D}{2}\zeta^{2}-\chi\right)  \cos\varphi_{\omega}-2D\zeta\left\vert
\alpha\right\vert \cos\varphi_{\alpha}\right]  dt^{\prime}\right)  \left\vert
\phi_{n}^{H}(t)\right\rangle .
\end{equation}

\section{ Few special examples}

\subsection{$\bigskip$Generalized time dependent non-Hermitian Swanson
Hamiltonian}

We now consider the SU(1,1) case first where $D=-2$. The SU(1,1) Lie algebra
has a realization in terms of boson creation and annihilation operators
$a^{+}$ and $a$ such that%
\begin{equation}
K_{0}=\frac{1}{2}\left(  a^{+}a+\frac{1}{2}\right)  ,\text{ \ \ }K_{-}%
=\frac{1}{2}a^{2},\text{ \ \ \ \ }K_{+}=\frac{1}{2}a^{+2}.
\end{equation}
\ \ \ \ \ \ 

When the Hamiltonian (\ref{HH}) is expressed in terms of position $x$ and
momentum $p,$ it describes the generalized quadratic time-dependent
non-Hermitian harmonic oscillator. The\ celebrated model of a non-Hermitian
PT-symmetric Hamiltonian quadratic in position and momentum was studied first
by Ahmed \cite{ahmed} and made popular by Swanson \cite{swanson} when
it\ was$\ $expressed in terms of the usual harmonic oscillator creation
$a^{+}$ \ and annihilation $a$ \ operators with $\omega,\alpha$ and $\beta$
time-independent real parameters, such that $\alpha\neq$ $\beta$ and
\ $\omega^{2}-4\alpha\beta>0$. This Hamiltonian has been studied extensively
in the literature by several authors
\cite{jones,bagchi,musumbu,quesne,sinha,eva}.

We construct here, by employing the Lewis-Riesenfeld method of invariants, the
solutions for the generalized version of the non-Hermitian Swanson Hamiltonian
with time-dependent coefficients\cite{fring2}%
\begin{equation}
H(t)=\omega(t)\left(  a^{+}a+\frac{1}{2}\right)  +\alpha(t)a^{2}%
+\beta(t)a^{+2},
\end{equation}
\ where $\left(  \omega(t),\alpha(t),\beta(t)\right)  $ $\in C$ are
time-dependent parameters. The form for $I^{PH}(t)$, which is both convenient
for calculations , is%
\begin{align}
I^{PH}(t)  &  =\exp\left[  \frac{\zeta}{2}a^{2}\right]  \exp\left[  -\frac
{\ln\vartheta_{0}}{2}\left(  a^{+}a+\frac{1}{2}\right)  \right]  \exp\left[
\frac{\zeta}{2}a^{+2}\right]  \text{ }\left(  a^{+}a+\frac{1}{2}\right)
\text{ }\nonumber\\
&  \times\exp\left[  -\frac{\zeta}{2}a^{+2}\right]  \exp\left[  \frac
{\ln\vartheta_{0}}{2}\left(  a^{+}a+\frac{1}{2}\right)  \right]  \exp\left[
-\frac{\zeta}{2}a^{2}\right]  ,
\end{align}
which brings out the Hamiltonian $2K_{0}=\left(  a^{+}a+\frac{1}{2}\right)  $
of the usual harmonic oscillator whose eigenstates $\left\vert n\right\rangle
$ ($n=0,1,3...$) and eigenvalues $\left(  n+\frac{1}{2}\right)  $ are well
known. As eigenstates of $I^{PH}(t)$ one can then take%

\begin{equation}
\left\vert \phi_{n}^{H}(t)\right\rangle =\exp\left[  \frac{\zeta}{2}%
a^{2}\right]  \exp\left[  -\frac{\ln\vartheta_{0}}{2}\left\{  \left(
a^{+}a+\frac{1}{2}\right)  -\left(  n+\frac{1}{2}\right)  \right\}  \right]
\exp\left[  \frac{\zeta}{2}a^{+2}\right]  \left\vert n\right\rangle ,
\end{equation}
the corresponding phase ( \ref{phase1} ) $\varphi_{n}(t)$ is
\begin{equation}
\varphi_{n}(t)=(n+\frac{1}{2})%
{\displaystyle\int\limits_{0}^{t}}
\frac{1}{\vartheta_{0}}\left[  \left(  \zeta^{2}+\chi\right)  \left\vert
\omega\right\vert \cos\varphi_{\omega}-4\zeta\left\vert \alpha\right\vert
\cos\varphi_{\alpha}\right]  dt^{\prime}.
\end{equation}

\subsection{A spinning particle in a time-varying magnetic field}

Now, we consider $D=2$ where the Hamiltonian (\ref{HH})\ and the invariant
(\ref{PH1}) possesse the symmetry of the dynamical group SU(2). There is
substantial literature on the time evolution of two-level system governed by a
non-Hermitian Hamiltonian \ $H(t)=\mathbf{B}(t)\mathbf{\sigma}$
\cite{1,2,3,4,5,6,7}, where $\mathbf{\sigma}$ is the vector of Pauli and the
components of the field $\mathbf{B}(t)$ are complex.

Knowing that, the ferromagnetic materials like Cobalt and Iron produce
magnetic fields whose magnitudes are measured by real numbers. Imaginary or
complex fields are, however, essential in the fundamental theory that
underlies the statistical physics of phase transitions, such as those
associated with the onset of magnetization. Long thought to be merely
mathematical constructs, a realization of these imaginary fields has now been
observed in magnetic resonance experiments performed on the spins of a
molecule \cite{8}, following an earlier theoretical proposal. A spin in a
time-varying complex magnetic field is a practical example for the case $D=2$
. Let
\[
K_{0}=J_{z},\ \ K_{-}=J_{-},\ \ \ \ K_{+}=J_{+},
\]
the Hamiltonian and the invariant are
\begin{align}
H(t)  &  =2\left[  \omega(t)J_{z}+\alpha(t)J_{-}+\beta(t)J_{+}\right]  ,\\
I^{PH}(t)  &  =\frac{2}{\vartheta_{0}}\left[  \left(  \zeta^{2}-\chi\right)
J_{z}-\chi\zeta J_{-}-\zeta J_{+}\right]  ,
\end{align}
where $\mathbf{J}$ is the spin angular momentum of the particle. The form for
$I^{PH}(t)$, which is both convenient for calculations , is%
\begin{equation}
I^{PH}(t)=\exp\left[  \zeta J_{-}\right]  \exp\left[  -\ln\vartheta_{0}%
J_{z}\right]  \exp\left[  \zeta J_{+}\right]  \text{ }J_{z}\text{ }\exp\left[
-\zeta J_{+}\right]  \exp\left[  \ln\vartheta_{0}J_{z}\right]  \exp\left[
-\zeta J_{-}\right]  .
\end{equation}
The instantaneous eigenstates of $I^{PH}(t)$ can be written in terms of the
eigenstates of $J_{z}$ denoted by $\left\vert m\right\rangle $, as%

\begin{equation}
\left\vert \phi_{m}^{H}(t)\right\rangle =\exp\left[  \zeta J_{-}\right]
\exp\left[  -\ln\vartheta_{0}\left(  J_{z}-m\right)  \right]  \exp\left[
\zeta J_{+}\right]  \left\vert m\right\rangle ,
\end{equation}
the corresponding eigenvalues are $m$. With the factor of $\exp\left[
m\ln\vartheta_{0}\right]  $ included in the definition of $\left\vert \phi
_{m}^{H}(t)\right\rangle $, the vector potential is singular only at the south pole.

For this case, the phase ( \ref{phase1} )$\varphi_{m}(t)$ is easy to calculate
and is given by
\begin{equation}
\varphi_{m}(t)=-m%
{\displaystyle\int\limits_{0}^{t}}
\frac{2}{\vartheta_{0}}\left[  \left(  \zeta^{2}-\chi\right)  \left\vert
\omega\right\vert \cos\varphi_{\omega}-4\zeta\left\vert \alpha\right\vert
\cos\varphi_{\alpha}\right]  dt^{\prime}.
\end{equation}
Before concluding this paper, we give a particular case when the parameters of
$H(t)$ are reals ; i.e., $\left(  \omega(t),\alpha(t),\beta(t)\right)  $ $\in
R$ .

\subsection{\-The Hamiltonian $H(t)$ with real coefficients $\omega
(t),\alpha(t),\beta(t)$}

When considering the time-dependent coefficients $\omega(t),\alpha
(t),\beta(t)$ to be real functions instead of complex ones, the polar angles
the polar angles $\varphi_{\omega}$, $\varphi_{\alpha}$, and $\varphi_{\beta}$
of $\omega$, $\alpha$, and $\beta$, vanish. By imposing that $\varphi_{\omega
}=\varphi_{\alpha}=$ $\varphi_{\beta}=0$, the Eqs.( \ref{cont1}- \ref{rel})
are simplified to%

\begin{equation}
\dot{\vartheta}_{0}=0,
\end{equation}

\begin{equation}
\overset{\cdot}{\zeta}=0,
\end{equation}

\begin{equation}
\
\begin{array}
[c]{c}%
\chi\left\vert \beta\right\vert =\left\vert \alpha\right\vert \\
\left(  \chi-\frac{D}{2}\zeta^{2}\right)  \left\vert \alpha\right\vert
=\chi\zeta\left\vert \omega\right\vert \\
\zeta\left\vert \omega\right\vert =\left(  \chi-\frac{D}{2}\zeta^{2}\right)
\left\vert \beta\right\vert
\end{array}
. \label{fin}%
\end{equation}
As one can see from the last equations that the metric parameters $\left(
\ref{TDC}\right)  \ \zeta$, $\vartheta_{0}$ are constants. Thus, the
time-dependent real coefficients $\omega(t),\alpha(t),\beta(t)$ of $H(t)$
provide a time-independent metric and consequently the gaugelike term
$i\hbar\dot{\eta}\left(  t\right)  \eta^{-1}\left(  t\right)  $ in the
quasi-Hermiticity relation $\left(  \ref{PHH1}\right)  $disappears and the
standard quasi-Hermiticity relation $\eta\left(  t\right)  H\left(  t\right)
=H^{\dag}\left(  t\right)  \eta\left(  t\right)  $ for\ the Hamiltonian $H(t)$
itself is recovered in complete analogy with the time-independent scenario.
Thus $H(t)$ is self-adjoined operator and therefore observable and can be
written in the following simple form%

\begin{equation}
H(t)=2\frac{\omega(t)}{\left(  \frac{D}{2}\zeta^{2}-\chi\right)  }\left\{
\left(  \frac{D}{2}\zeta^{2}-\chi\right)  K_{0}-\chi\zeta K_{-}-\zeta
K_{+}\right\}  =\frac{\omega(t)\vartheta_{0}}{\left(  \frac{D}{2}\zeta
^{2}-\chi\right)  }I^{PH}(t),
\end{equation}
which reveal its self-adjoined character and therefore its observability. From
Eqs.( \ref{fin}), we derive the metric parameter $\zeta$ in terms of
parameters of the Hamiltonian $H(t)$
\[
\zeta=\frac{1}{2\left\vert \beta\right\vert }\left(  -\frac{D}{2}\left\vert
\omega\right\vert \pm\sqrt{\left\vert \omega\right\vert ^{2}+2D\left\vert
\alpha\right\vert \left\vert \beta\right\vert }\right)  .
\]

\section{Conclusion}

The results we have presented offer a general and comprehensive treatment of
the non-Hermitian dynamics of SU(1,1) and SU(2) quantum systems. Non-Hermitian
Hamiltonian operators have been the subject of considerable interest during
the last years within the framework of the PT symmetry and pseudo-Hermiticity theories.

Recently, It has demonstrated that a time-dependent metric operator cannot
ensure the unitarity of the time evolution simultaneously with the
observability of the Hamiltonian and thus the general framework for a
description of a time evolution for time-dependent non-Hermitian Hamiltonians
has been stated \cite{fring1,fring2}. A well-known method based on a
time-dependent unitary transformation for the treatment of time-dependent
Hermitian Hamiltonians \cite{mus1,mizrahi}, has been adapted by the authors of
Ref. \cite{fring2} to solve the the time-dependent Dyson and the
time-dependent quasi-Hermiticity relations for non-Hermitian Swanson
Hamiltonian with time-dependent coefficients, where the time-dependent unitary
transformation is replaced by a non-unitary transformation to conform to
non-Hermitian Hamiltonians.

In this work, using the standard quasi-Hermiticity relation (\ref{quasi})
between a non-Hermitian invariant operator $I^{PH}(t)$ and a Hermitian one
$I^{h}(t)$, we have considered the dynamical behavior of SU(1,1) and SU(2)
non-Hermitian time-dependent quantum systems by presenting an alternative
approach to solve it. We investigated in detail the main frames of
time-dependent non-Hermitian SU(1,1) and SU(2) systems in the framwork of the
Lewis and Riesenfeld method which ensures that a solution of the
Schr\"{o}dinger equation governed by a time-dependent non-Hermitian
Hamiltonian is an eigenstate of an associated pseudo-Hermitian invariant
operator $I^{PH}(t)$ with a time-dependent global real phase factor
$\varphi_{n}(t)$.

The properties derived here help us to understand\ better\ systems described
by time-dependent non-Hermitian Hamiltonians and should play a central role in
time-dependent non-Hermitian quantum mechanics. After going through these
properties, we then have presented two illustrative examples: the generalized
Swanson model \ and a spinning particle in a time-varying magnetic field. When
a time dependent parameters $\omega(t),\alpha(t),\beta(t)$ are supposed real,
the standard quasi-Hermiticity relation for the time dependent Hamiltonian
$H\left(  t\right)  $ occur and the metric operator become time-independent.

\end{document}